\documentclass[10pt,letterpaper]{article}
\usepackage{opex3}
\usepackage{color}
\usepackage{graphicx}
\usepackage{subfig}
\begin{document}

\title{A Simple and Efficient Absorption Filter for Single Photons from a Cold Atom Quantum Memory}

\author{Daniel T. Stack, Patricia J. Lee, and Qudsia Quraishi}

\address{Quantum Sciences Group, Army Research Laboratory, Adelphi, Maryland 20783, USA}

\email{qudsia.quraishi.civ@mail.mil}

\begin{abstract*}
The ability to filter unwanted light signals is critical to the operation
of quantum memories based on neutral atom ensembles. Here we demonstrate an efficient frequency
filter which uses a vapor cell filled with $^{85}$Rb and a buffer
gas to attenuate both residual laser light and noise photons by nearly two orders of magnitude with little
loss to the single photons associated with our cold
$^{87}$Rb quantum memory. This simple, passive filter provides an
additional 18 dB attenuation of our pump laser and erroneous spontaneous emissions for every 1 dB loss
of the single photon signal. We show that the addition of a
frequency filter increases the non-classical correlations and readout
efficiency of our quantum memory by $\approx35\%$.
\end{abstract*}


\section{\label{sec:intro} Introduction}

Distribution of entangled quantum states over significant distances
is important to the development of future quantum technologies such
as long-distance cryptography, networks of atomic clocks, distributed
quantum computing, etc.~\cite{bennet_c._h._quantum_1984,gisin_quantum_2002,kimble_quantum_2008,nielsen_quantum_2013,komar_quantum_2014}.
Quantum repeaters can overcome the problems of exponential losses
in direct transmission of single photons over quantum channels through
entanglement swapping of remotely placed quantum memories via single photon detection~\cite{briegel_quantum_1998}.
Long-lived quantum memories and single photon sources are building
blocks for systems capable of efficiently and securely transferring
quantum information over extremely long distances.  Quantum memories are a very active area of research with many different quantum systems as viable candidates, including ions~\cite{olmschenk_quantum_2010}, neutral atom ensembles~\cite{matsukevich_quantum_2004}, Rydberg atoms~\cite{wilk_entanglement_2010,li_entanglement_2013,zhao_efficient_2010}, single neutral atoms~\cite{nolleke_efficient_2013}, quantum dots~\cite{kroutvar_optically_2004}, and nitrogen vacancy centers~\cite{lee_entangling_2011}. Cold ensembles
of neutral atoms are a favorable platform for quantum memories and
single photon sources as they can store single excitations in long-lived ground states that lead to
extremely long coherence times \cite{kuzmich_generation_2003,dudin_light_2013}.

The ability to store and retrieve single excitations efficiently and
background-free is crucial to building quantum memories based on neutral
atom ensembles~\cite{duan_long-distance_2001}. The generation (retrieval)
of these excitations usually requires a strong laser beam to create
a coherent excitation in the atomic ensemble that can be heralded
(read out) by the detection of a single photon from the ensemble.
The detection of classical pump photons at the single photon detector
is a source of error in standard quantum communication protocols that
needs to be eliminated. This strong pump can be filtered by a variety
of means: spatial, polarization, frequency, etc. Spatial filtering
is achieved using the off-axis geometry pioneered in \cite{braje_frequency_2004}.
By aligning the single photon arms a few degrees off-axis from the pump arms, noise suppression $>40$ dB is possible, usually limited by scattering off the surfaces near the memory. 
Polarization filtering is accomplished by post-selecting single
photons that are orthogonally polarized to the corresponding pump
fields using properly aligned waveplates and polarizing optical elements. Birefringence of optical elements and imperfect polarizers usually limit noise suppression to $\sim 40$ dB as well~\cite{kupchak_room-temperature_2014}.
Optical cavities \cite{nolleke_efficient_2013}, optically pumped atomic
vapor cells \cite{laurat_efficient_2006}, and fiber Bragg gratings \cite{fernandez-gonzalvo_quantum_2013}
are just a few examples of frequency selective elements that have
been used in various quantum memory experiments to reject sources
of noise. It is possible to achieve extremely high levels of noise rejection ($>100$ dB) though usually at the cost of lower transmission probability ($\approx~10\%$)~\cite{kupchak_room-temperature_2014}. Each of these frequency filters require complicated electronics, significant temperature stabilization, external references, or additional lasers.

\begin{figure*}[t]
\centering
\includegraphics[width=0.65\columnwidth]{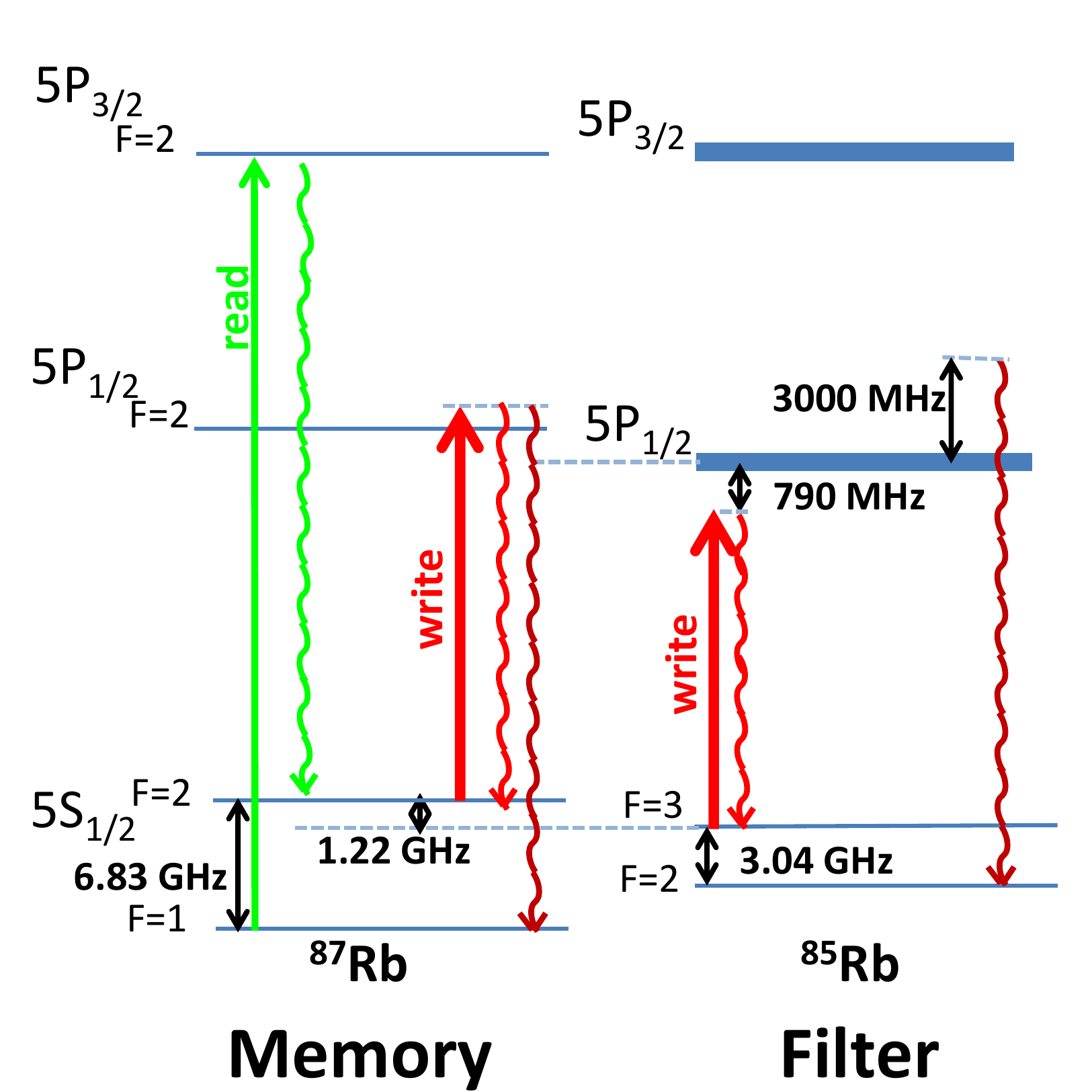}
\protect\protect\caption{\label{fig:levels} The relevant energy levels and transitions of $^{87}$Rb
and $^{85}$Rb used in this paper. The write beam is 20, 40, or 60 MHz
blue-detuned from the $5S_{1/2},F=2\rightarrow 5P_{1/2},F=2$ (D1)
transition. Atoms excited to the $5P_{1/2},F=2$ state can decay to
either the $5S_{1/2},F=1$ or the $5S_{1/2},F=2$ ground state in $^{87}$Rb.
The write beam and the emission from the $5P_{1/2},F=2\rightarrow5S_{1/2},F=2$
state are detuned from the Doppler-broadened $5S_{1/2},F=3\rightarrow5P_{1/2}$ transition
in $^{85}$Rb by $\approx$ 790 MHz. The emission from the $5P_{1/2},F=2\rightarrow5S_{1/2},F=1$
state is detuned from the Doppler-broadened $5S_{1/2},F=2\rightarrow5P_{1/2}$ transition
in $^{85}$Rb by $\approx$ 3000 MHz. This difference in the detuning
of these transitions serves as the basis for using a $^{85}$Rb vapor
cell as a frequency filter in our quantum memory experiment. The read
beam is resonant with the $5S_{1/2},F=1\rightarrow5P_{3/2},F=2$ (D2) transition
in $^{87}$Rb. Drawing not to scale.}
\end{figure*}

In this paper, we show how the serendipitous near degeneracy of spectral lines in $^{85}$Rb and $^{87}$Rb allow one to construct an efficient frequency
filter for both the classical  pump beams and erroneous atomic decays in a neutral atom quantum memory,
without strong filtering of the desired signal that herald the creation
of these coherent excitations. To accomplish this we use the two naturally-occurring
isotope of rubidium, $^{87}$Rb for use as a quantum memory and a vapor cell filled with isotopically-pure $^{85}$Rb
to filter the pump photons associated with this process (Fig.~\ref{fig:levels}).
By properly choosing the correct transition in $^{87}$Rb for the
creation of spin-waves, the frequency of the Write beam and Noise single photons are nearly
resonant with a transition in $^{85}$Rb, while the single photons
that are correlated to the created spin-waves are much further detuned from
any transition in $^{85}$Rb. The Noise single photons arise from spontaneous decays back to the original ground state ($5S_{1/2},F = 2$). This method for filtering radiation
is well known in the atomic clock community \cite{camparo_rubidium_2007},
and has been used in warm vapor quantum memory experiments previously \cite{heifetz_super_2004,manz_collisional_2007,bashkansky_quantum_2012}. 
A vapor cell frequency filter would be much more effective for a cold atom quantum memory because the Noise and Signal emissions from the cold ensemble are limited only by the atomic linewidth and not the much larger Doppler-broadened linewidth for warm vapor memories. A broader linewidth makes it more difficult to efficiently filter Noise processes while not absorbing the single photons heralding the creation of spin-waves needed for the operation of the quantum memory.

First, we build a simple model to predict the utility of such a filter
under conditions that can be verified experimentally. Then, we perform
an experiment where the filter is used to block spontaneously emitted
photons not associated with the creation of spin-waves, which would
contribute to a reduced readout efficiency. The result is a quantum memory with higher readout efficiency, hence 
higher single photon rates and greater nonclassical correlations.

\section{\label{sec:theory} Theory}

\begin{figure*}[t]
\centering
\subfloat[]{\includegraphics[width=0.45\linewidth]{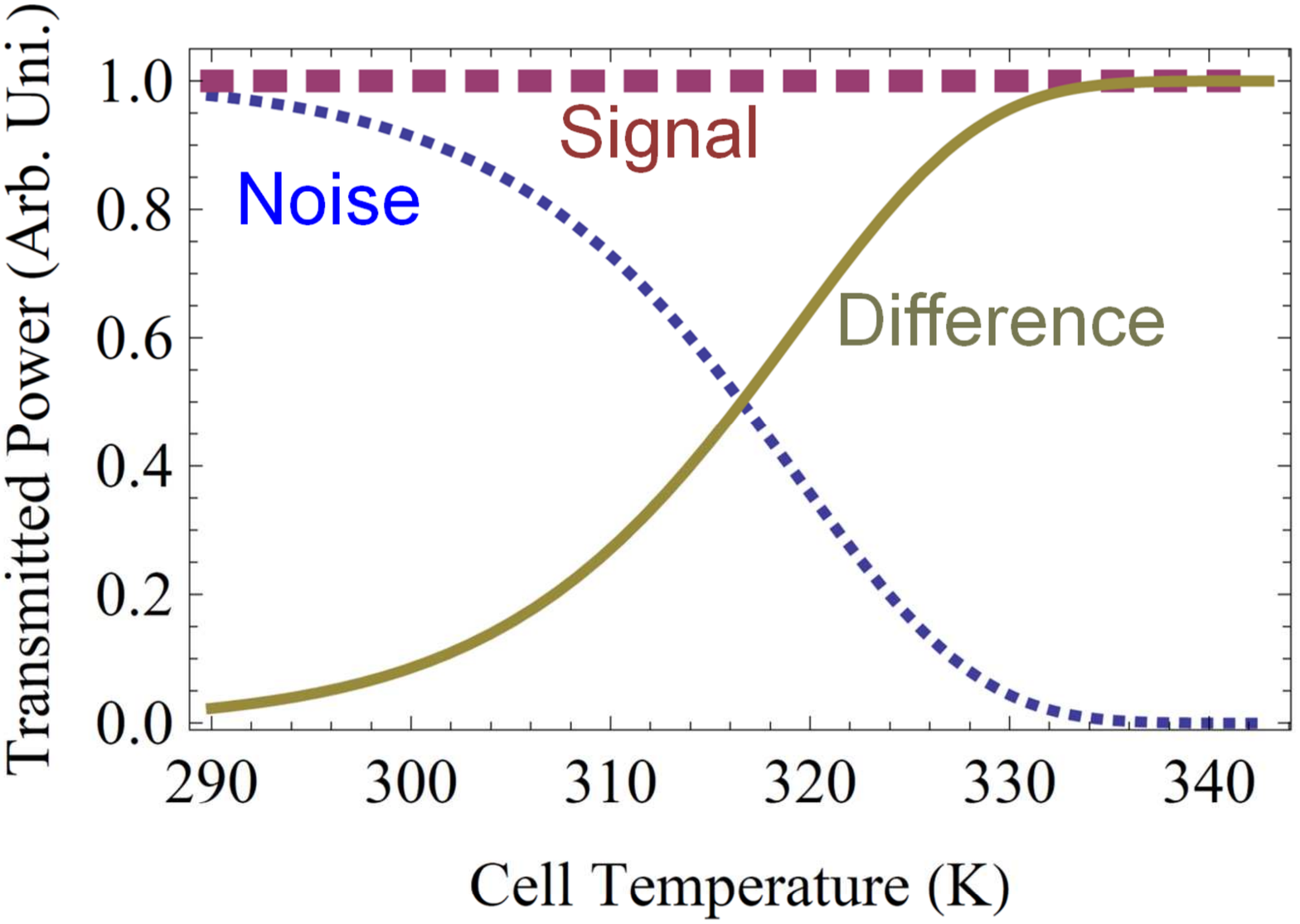}}
\subfloat[]{\includegraphics[width=0.45\linewidth]{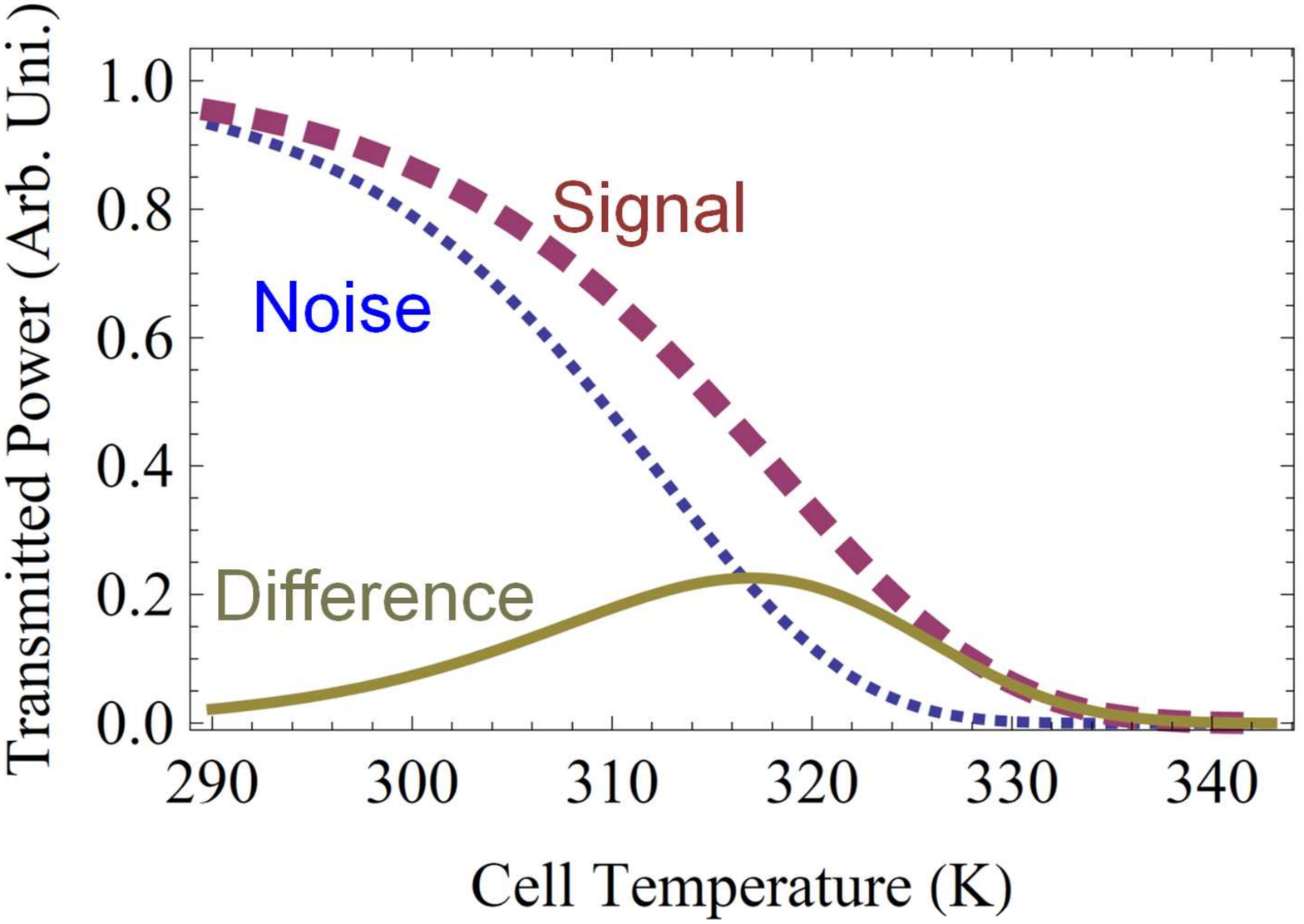}}\protect\caption{\label{fig:dopp}
Plots of the transmitted light through the vapor
cell filter as a function of temperature in the Doppler-broadened
limit (Eq. \ref{eq:dopp}) for a filter cell length of 1 inch. \textbf{(a)} The pure $^{85}$Rb filter cell acts nearly perfectly by attenuating
the undesired light strongly (blue, dotted) and no attenuation of the signal
(magenta, dashed) leading to large difference in attenuation between the two
signals (gold, solid). \textbf{(b)} Small concentrations of $^{87}$Rb
($^{87}$Rb$/^{85}$Rb fraction of $0.2\%$) result in strong attenuation
of the resonant background and signal fields. The emission from the
cold $^{87}$Rb ensemble is assumed to be hyperfine specific but the
absorption by the warm $^{85}$Rb vapor cell transitions are assumed
to be Doppler-broadened and not hyperfine specific (See Fig.~\ref{fig:levels}). }
\end{figure*}

To construct a frequency filter that is capable of attenuating the
undesired background radiation it will be necessary to calculate the
strength and the lineshape of the filter as a function of vapor cell
temperature and buffer gas pressure. In the limit that the frequency of
the input light is much more narrow-band (roughly the atomic linewidth
$\Gamma/2\pi \approx 5.8$ MHz) than the bandwidth of the frequency filter ($\gg 100$ MHz)~\cite{milonni_laser_2010},

\begin{equation}
\frac{I_{OUT}(z)}{I_{IN}}=e^{-a(\delta,T,P)z}\label{eq:att}
\end{equation}
where $a(\delta,T,P)$is the attenuation per unit length that depends
on the detuning from the nearest resonance line $\delta,$ temperature
of the filter cell $T$, and buffer gas pressure $P$, while $z$
is the length of the filter cell. The lineshape of the filter function
may depend on both the inhomogeneous broadening due to the Doppler effect
of atoms at a temperature $T$ with a characteristic FWHM of $\delta v_{D}(T)$
and the homogeneous broadening due to collisions with a buffer gas
at pressure $P$ with a characteristic FWHM of $\gamma(P)$. The
vapor cell filter is considered in two limits: when Doppler effects dominate and when atomic collisions dominate.

In the Doppler-broadened case:
\begin{equation}
a(T)=f(T)\exp[\frac{-4\ln2\delta_{i}^{2}}{(\delta v_{D}(T))^{2}}]\label{eq:dopp}
\end{equation}
where $f(T)$ is proportional to the vapor cell atom number and increases
exponentially with cell temperature~\cite{steck2001rubidium}, $\delta v_{D}(T)$ is the Doppler-broadened
FWHM of the $^{85}$Rb lines, and $\delta_{i}$ is either the
Noise or Signal frequency detuning from the nearest $^{85}$Rb resonance shown in Fig.~\ref{fig:levels}(a).

One would like a simple/passive frequency filter for strong attenuation
of unwanted background radiation with weak attenuation of
the signal photons associated with the created spin waves. Figure~\ref{fig:dopp}(a) shows that an optically
thick vapor cell of pure $^{85}$Rb with no buffer gas would accomplish
this due to the exponential falloff of the filter strength as a function
of $\delta^{2}$ and the detuning of the light associated with
the desired transition is roughly four times further detuned than the
light associated with the undesired transition (790 MHz vs 3000 MHz). However, contamination of the
cell with a small amount $^{87}$Rb can have a significant effect on the attenuation of the Signal photons.
If the ratio of $^{87}$Rb$/^{85}$Rb is on the order of a few times
$10^{-3}$ (which is roughly the level of purification available commercially)
this will result in roughly equal attenuation of the Noise photons
by $^{85}$Rb and Signal photons by $^{87}$Rb, negating the utility
of the frequency filter in Figure~\ref{fig:dopp}(b). It is therefore
prudent to consider the frequency filter in the presence of a buffer
gas to see if the effect of $^{87}$Rb can be mitigated.

\begin{figure}
\centering\includegraphics[width=0.6\columnwidth]{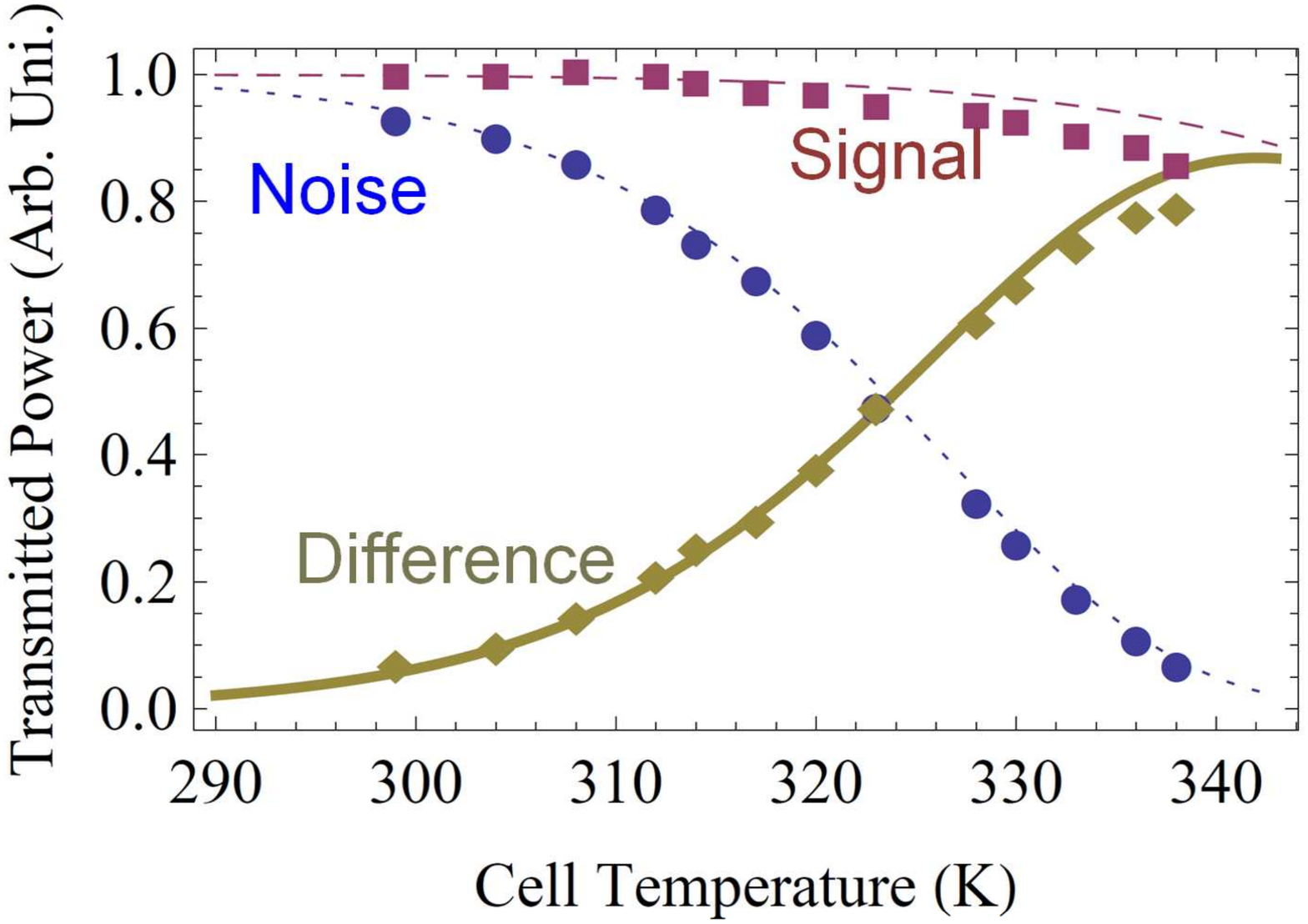}
\protect\caption{\label{fig:pres}
Normalized measured transmission
of classical laser light at the Noise (circles) and Signal (squares)
frequencies and the difference between the two signals (diamonds)
at a variety of vapor cell temperatures with 47 Torr of Ar in a 1 inch long $^{85}$Rb vapor cell. The corresponding
solid lines are calculations of the transmitted light through the
vapor cell filter as a function of temperature with 47 Torr of Ar
using the filter response from Eq. \ref{eq:pres}.}
\end{figure}

In the pressure-broadened case:
\begin{equation}
a(T,P)=f(T)\frac{\alpha P}{\pi(\delta_{i}-\beta P)^{2}-(\frac{\alpha P}{2})^{2}}\label{eq:pres}
\end{equation}
where $\beta$ is the coefficient for the linear frequency shift with
a value of $-6.7$~MHz/Torr and $\alpha$ is the linear coefficient
for the pressure-broadened linewidth with a value of $21$ MHz/Torr
at $T=300$ K for the D1 transition due to collisions of Rb with
an Argon buffer gas. Since we will only vary the temperature of the
vapor cell by $\sim10\,\%$ it will be a good approximation to say
that these pressure-dependent coefficients are constant as a function of temperature. 
Figure~\ref{fig:pres} is a plot of Eq.~\ref{eq:att} with $a(T,P)$ from Eq.~\ref{eq:pres} which shows how the introduction of 47 Torr of an Ar buffer
gas changes the filter response as a function of vapor cell temperature. The Ar pressure of 47 Torr was chosen such that the filter could be modeled in the pressure-broadened limit ($\alpha P > \delta \nu_D$) where the filter is most effective. Increasing the buffer gas pressure significantly would dilute the filter strength by broadening the filter response function. From Fig.~\ref{fig:pres} one can extract an 18 dB attenuation from the Noise for every 1 dB attenuation of the Signal over the range of filter cell temperatures measured. Disagreements between data and theory at higher filter cell 
temperatures may be explained by either violations of assumptions made in our simple model 
(such as temperature independent $\gamma(P)$) or greater $^{87}$Rb contamination in the vapor cell than 
the specification from the manufacturer ($^{87}$Rb$/^{85}$Rb~$\approx~0.2\%$).

The optimum operating temperature for the vapor cell filter should correspond to roughly the maximum difference between the Signal and Noise if one assumes roughly equal contributions of Signal and Noise before the frequency filter. This strikes a balance between increasing the Readout Efficiency and $g^{(2)}$ correlations (strong attenuation of Noise) while maximizing the coincident count rate (weak attenuation of Signal). 
With a buffer gas of sufficient pressure and high enough operating temperature it is possible to achieve
significant attenuation of the background light with only a small
amount of signal loss.

\section{\label{sec:exp} Experimental Results and Methods}

\subsection{\label{subsec:mem} Neutral Atom Quantum Memory}

\begin{figure*}
\centering
\captionsetup[subfigure]{labelformat=empty}
\subfloat[]{\includegraphics[width=0.75\columnwidth]{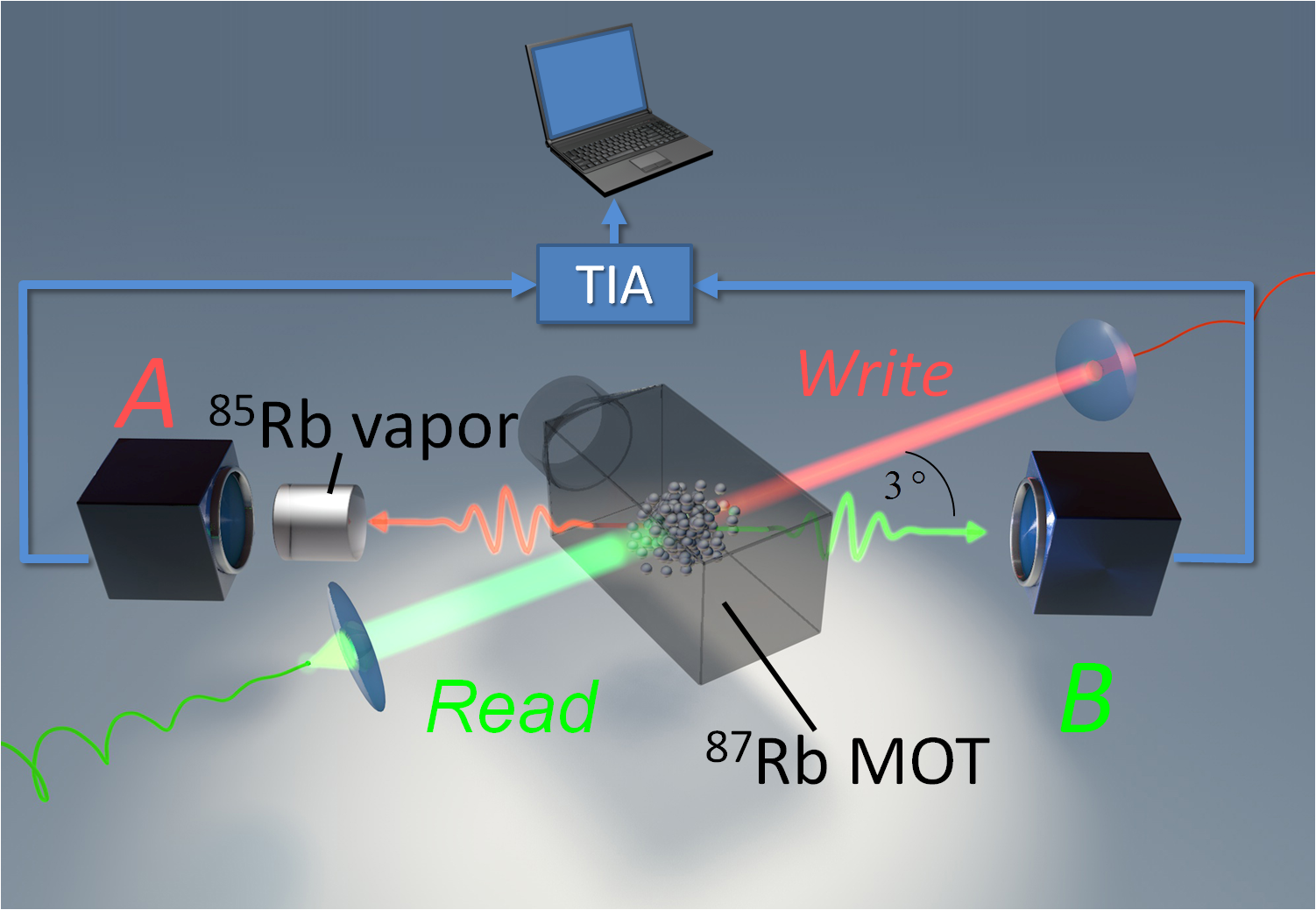}}
\llap{
 \LARGE \parbox[b]{0.7in}{(\textbf{a})\\\rule{0ex}{0.0in}
  }}
\subfloat[]{\includegraphics[width=.75\columnwidth]{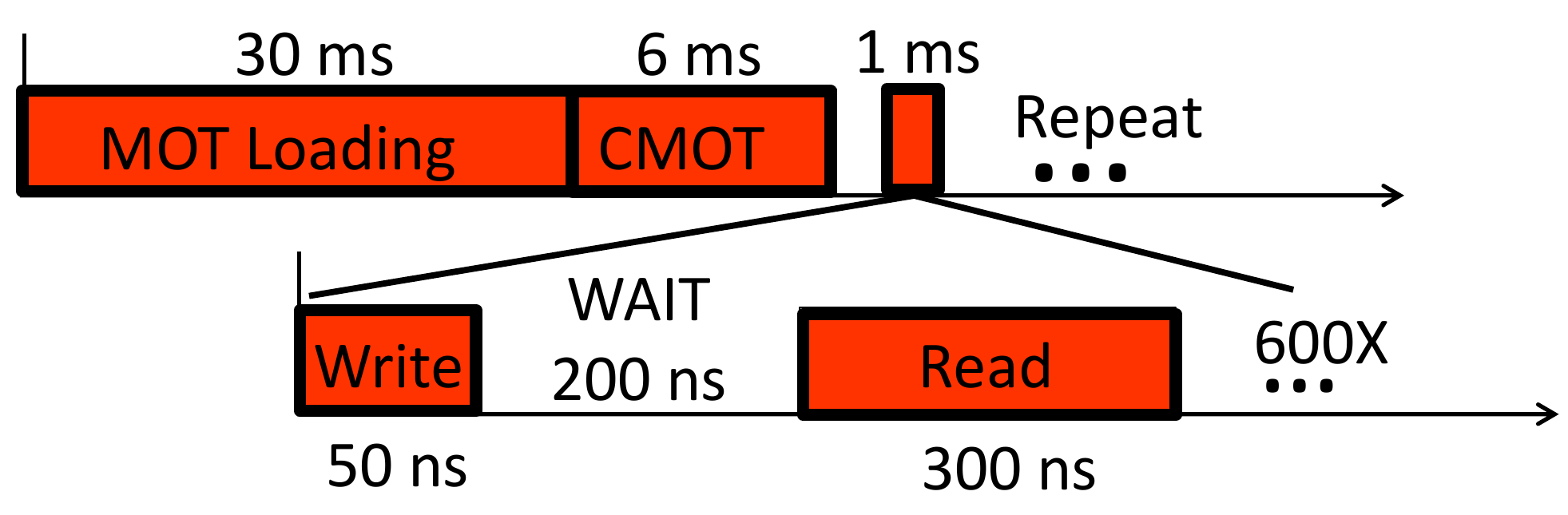}}\llap{
  \LARGE \parbox[b]{0.7in}{(\textbf{b})\\\rule{0ex}{1.05in}
  }}
\subfloat[]{\includegraphics[width=.75\columnwidth]{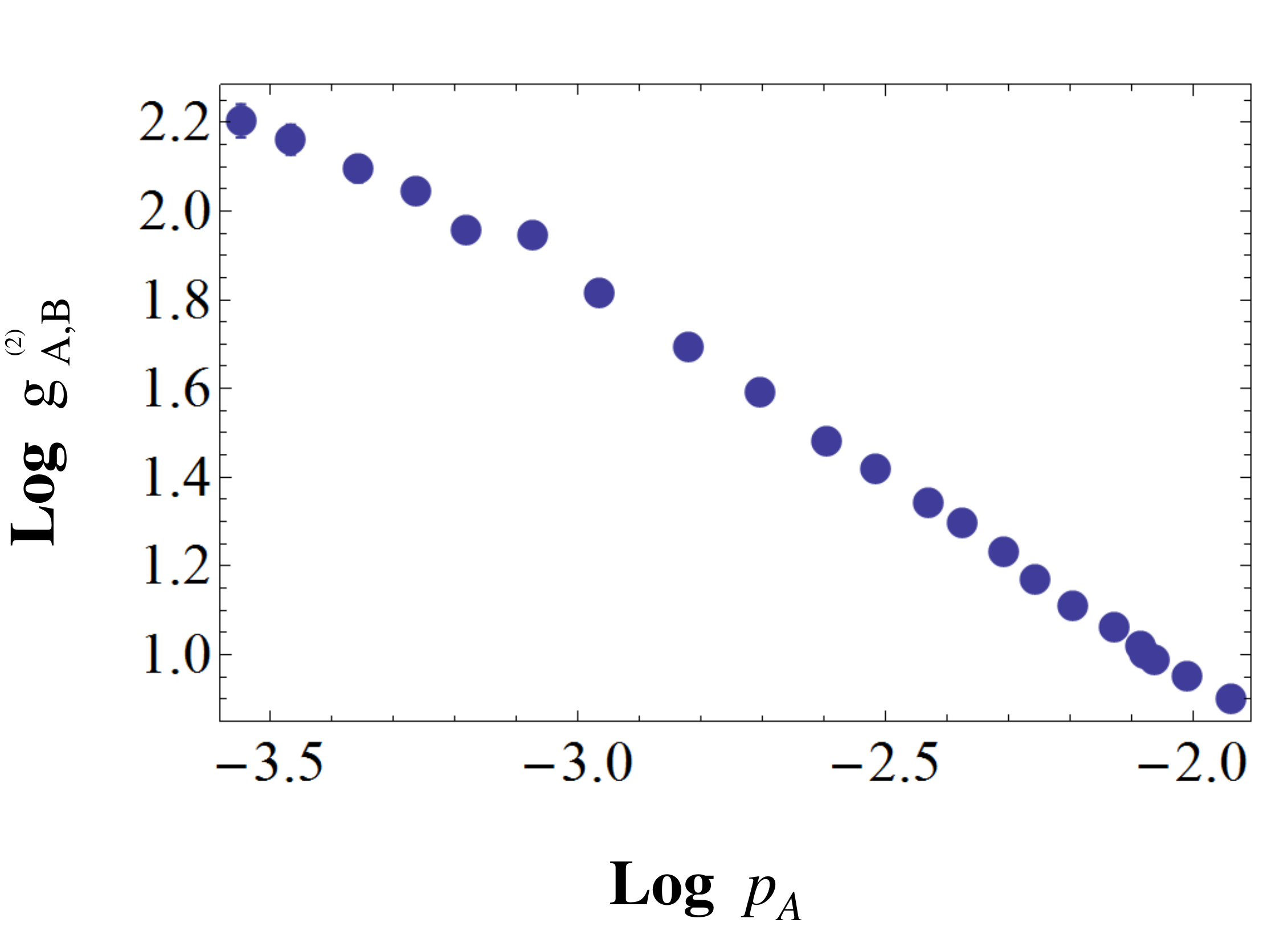}}\llap{
 \LARGE \parbox[b]{0.7in}{(\textbf{c})\\\rule{0ex}{2.0in}
  }}
\protect\caption{\label{fig:g2} \textbf{(a)} The experimental setup for our quantum memory. Counter-propagating Write and Read light pulses interact with a cold $^{87}$Rb ensemble while emitted single photons are detected off-axis by Si APDs. Arrival times of the photons are measured by a time-interval analyzer (TIA) and sent to a PC for analysis. Emitted Write photons are filtered by a warm $^{85}$Rb vapor cell to attenuate Noise photons.
\textbf{(b)} Timing sequence for the production of a cold ensemble and interaction with Write and Read beams. 
A MOT is loaded and compressed to create a cold, dense atomic medium at a rate of 27 Hz.  
The MOT magnetic and laser fields are then extinguished for a 1 ms duration and a series of 600 Read-Write pulses interact with the system.
\textbf{(c)} Log-Log plot of $g_{A,B}^2$ as a function of $p_A$ with filter cell at a temperature of $336~K$ and $\Delta_w/2\pi=20$ MHz.  Write probability ($p_A$) is varied from $10^{-2}-10^{-4}$ and the corresponding values of $g_{A,B}^{(2)}$ increase from $\approx 8$ to $> 160$.}
\end{figure*}

A magneto-optical trap (MOT) of $^{87}$Rb atoms produces an optically
thick atomic ensemble for our experiment (Fig.~\ref{fig:g2}(a)). Approximately 25 mW/cm$^2$ of cooling
light, detuned 15 MHz below the $5S_{1/2},F=2\rightarrow5P_{3/2},F=3$
transition in $^{87}$Rb, and 5 mW of repump light, on resonance with
the $5S_{1/2},F=1\rightarrow5P_{3/2},F=2$ transition, are delivered
collinearly to the vacuum system. Loading from a room-temperature
Rb vapor, a cloud of $\sim1.8\times10^{8}$ atoms at a temperature
of $\sim$100 $\mu$K is produced in roughly 5 seconds with a magnetic field gradient of 10 G$/$cm 
(subsequent reloading of the MOT is performed in 30 ms as seen in Fig.~\ref{fig:g2}(b)).  This is then followed by a compressed MOT stage that consists of simultaneously lowering the intensity of the repump and
increasing the gradient of the quadrupole magnetic field to 22 G$/$cm for 6 ms before turning the
field off. The cooling light is turned off 100 $\mu$s later, and
the repump light 2 $\mu$s after that, leaving all of the atoms in
the $5S_{1/2},F=2$ ground state.

A sequence of 1000 trials with individual durations of $\sim1\,\mu s$
is performed as follows: a weak, linearly-polarized, Write pulse, tuned 20, 40, or 60 MHz above the
$5S_{1/2},F=2\rightarrow 5P_{1/2},F=2$ (See Fig.~\ref{fig:levels}) transition is shone on the atomic ensemble. The Write pulse has a 400 $\mu m$ waist and 50 ns duration set by an acousto-optical modulator and an electro-optic modulator. For low enough
excitation probability ($10^{-2}-10^{-4}$) a photon detection within an 80 ns time interval at Si avalanche photodiode (APD) A heralds the transfer of one $^{87}$Rb atom in the ensemble from the $F=2$ ground state
to the $F=1$ ground state in a particular spatial mode that has a waist of 200 $\mu$m at the MOT. From measurements of the atomic density, beam size, and length of the ensemble we can estimate that $\approx$ 10$^6$-10$^7$ atoms participate in this process. The polarization of the generated single
photon is post-selected to be orthogonal to the Write pulse with appropriate waveplates and linear polarizers. 
After a hold time of 200 ns, a strong, linearly-polarized (orthogonal to the Write), resonant Read pulse of duration 300 ns and 400 $\mu m$ waist interacts with
the ensemble, producing a photon that is detected
by APD B within a 100 ns window. The generated
photon has a polarization orthogonal to the Read pulse. The single photon detection arms are aligned at an angle of $3^{\circ}$ relative to the classical pump beams. By aligning the single photon arms a few degrees off-axis from the pump arms, the pump light is suppressed by a measured value of $\approx 40$ dB in the single photon detection arms. The APD signals
are sent to a time-interval analyzer (TIA) for data collection and analysis.

The strength of the correlations between the photon-pairs generated
by the atomic ensemble can be measured by the normalized intensity
cross-correlation function: 
\begin{equation}
g_{A,B}^{(2)}=\frac{p_{AB}}{p_{A}p_{B}}=\frac{RE}{p_B}
\label{eq:g2}
\end{equation}
where $p_{AB}$ is the probability to detect a photon pair, $p_{i}$
is the probability for an event click at detector $i$, and $RE$ is the readout efficiency defined as $p_{AB}/p_A$.
A plot of $g_{A,B}^{(2)}$ over a range of different write probabilities, $p_A$, is shown in Figure~\ref{fig:g2}(c).  
Figure~\ref{fig:g2}(c) shows a series of measurement of $g_{A,B}^{(2)}$ over a range of values for $p_A$ ($10^{-2}-10^{-4}$) 
with $\Delta_w/2\pi=20$ MHz and a constant filter cell temperature of $336~K$. The value of $g_{A,B}^2$ saturates at low $p_A$ due to dark counts in the APDs and contamination of the readout arm by classical Read light. The transmission values for the Noise and Signal photons 
in Fig.~\ref{fig:pres} at a filter cell temperature of $336~K$ are 12$\%$ and 88$\%$ respectively. The filter cell attenuates the 
Noise photons significantly which increases the measured $RE$ and $g_{A,B}^{(2)}$ for a given $p_A$ while not significantly 
attenuating the Signal photons. Attenuation of the Signal photons would lead directly to a decrease in the coincidence probability, 
$p_{AB}$, an undesirable outcome for entanglement creation and entanglement swapping operations.

\subsection {\label{subsec:sin} Filtering Single Photon Noise}

\begin{figure}
\centering
\subfloat[]{\includegraphics[width=0.55\columnwidth]{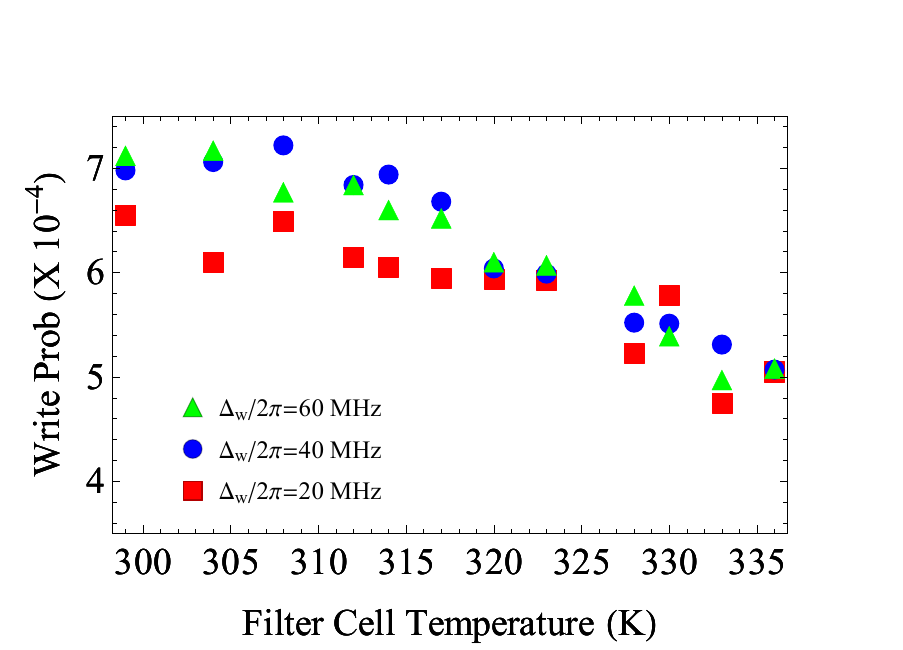}}
\subfloat[]{\includegraphics[width=0.55\columnwidth]{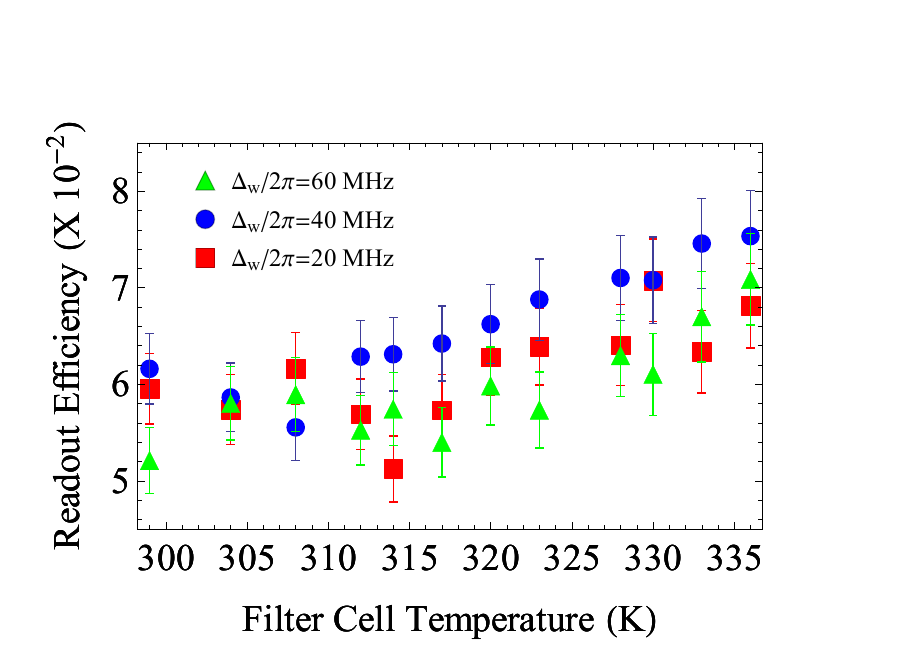}}
\protect\caption{\label{fig:wrre}Probability of detecting a write photon and Readout Efficiency as
a function of the filter cell temperature for $\Delta_w/2\pi = $20, 40, and 60~MHz. \textbf{(a)} The 
write probability decreases by $\approx 40\%$ over the range of filter cell temperatures used in this experiment for each write beam detuning.
\textbf{(b)} A significant increase in Readout Efficiency as the filter cell temperature is increased is observed. 
The decrease in Noise photon detection in the Write arm leads to an $\approx 35\%$ increase in the Readout Efficiency for $\Delta_w/2\pi=$ 40 and 60~MHz and $\approx 15\%$ for $\Delta_w/2\pi=20$ MHz. 
 Each point in plots (a) and (b) correspond to $\sim 10000$ write detection events and $\sim 500$ coincident counts respectively. 
The running time required to obtain a single data point is $\approx 12$ minutes. Error bars are calculated assuming Poissonian statistics.}
\end{figure}

In this section we study the effect of a $^{85}$Rb filter on the operation of a cold $^{87}$Rb quantum memory.  
This was done by varying the temperature of AR-coated vapor cell filled with isotopically-pure $^{85}$Rb 
and 47 Torr of Argon from $298~K$ to $336~K$. 
Based upon Fig.~\ref{fig:pres} we would expect the presence of $^{85}$Rb vapor cell filter to have only 
a small effect on the quantum memory parameters ($p_A, \;RE, \;g_{A,B}^{(2)}$) at an operating temperature of $298\;K$. At this temperature 
the filter cell absorbs only $10\%$ of the Noise photons and less than $1\%$ of the Signal photons.
 However, as the filter cell temperature is increased, an increasing percentage of Noise photons (pump and emitted) are filtered
 by the cell with corresponding decreases in $p_A$ and increases in $RE$ and $g_{A,B}^{(2)}$. 
Fig.~\ref{fig:wrre}(a) shows the decrease in write probability as a function of filter cell temperature. For $\Delta_w/2\pi= 20$ MHz the Write probability decreases from a value
of $6.5 \times 10^{-4}$ to $5 \times 10^{-4}$ as the filter cell temperature is increased from $298~K$ to $336~K$. For Write detunings of 40 MHz and 60 MHz the Write probability  decreases from $7 \times 10^{-4}$ to $5 \times 10^{-4}$ over the same temperature range.
Most of this decrease can be attributed to the vapor cell filter absorption of Noise photons emitted by the cold ensemble taking into account the results of Fig.~\ref{fig:pres}. At a filter cell temperature of $336~K$ only 6$\%$ of Noise photons are transmitted through the vapor cell while $> 85\%$ of Signal photons are transmitted. 
The vapor cell also filters out pump photons from the Write beam that scatter off of the vacuum system glass walls. 
However the effect on the quantum memory parameters is minimal because the pump contamination in the Write detection arm is only $\sim 1 \times10^{-5}$. 
Since the filter cell has little effect on the coincident counts between detectors A and B, there is then 
a corresponding increase of $\approx 35\%$ in the measured Readout Efficiency as seen Fig.~\ref{fig:wrre}(b) 
for $\Delta_w/2\pi=$ 40 and 60~MHz and $\approx 15\%$ for $\Delta_w/2\pi=20$ MHz. Readout Efficiency increases from 6$\%$ to 7$\%$ for a Write detuning of 20 MHz, from 5.8$\%$ to 7.5$\%$ for a Write detuning of 40 MHz, and 5.5$\%$ to 7$\%$ for a Write detuning of 60 MHz.

\begin{figure}
\centering
\includegraphics[width=0.9\columnwidth]{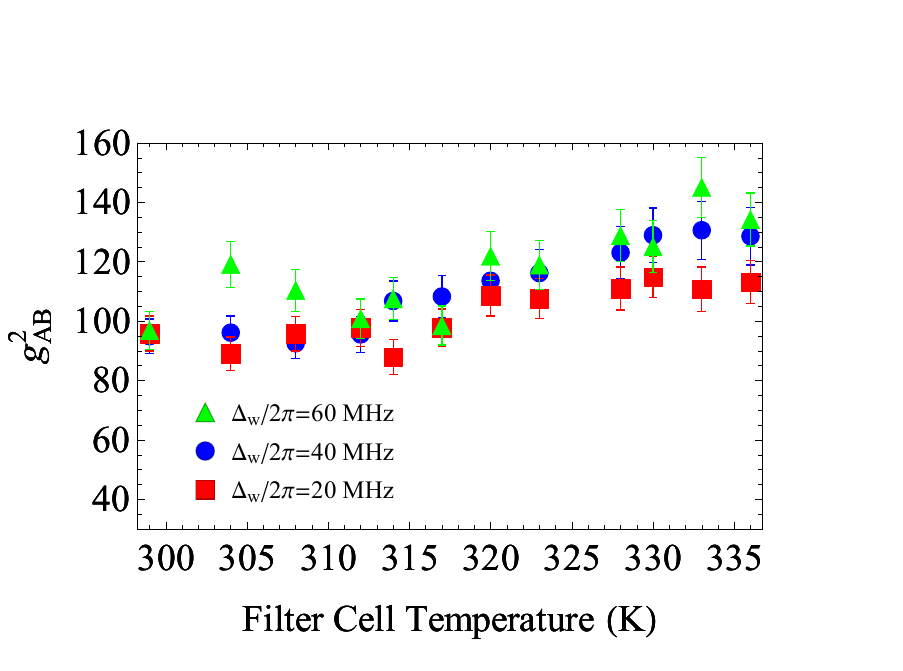}
\protect\caption{\label{fig:g2freq}$g_{A,B}^{(2)}$ correlations as a function of filter cell temperature for $\Delta_w/2\pi=$20, 40 and 60 MHz. 
The strength of the correlations increase by $\approx 15\%$ for $\Delta_w/2\pi = 20$ MHz,
 $\approx 35\%$ for $\Delta_w/2\pi = 40$ MHz, and $\approx 35\%$ for $\Delta_w/2\pi = 40$ MHz. The increase in non-classical correlations follows directly from the
 increase in Readout Efficiency seen in Fig.~\ref{fig:wrre}. Error bars are calculated assuming Poissonian statistics.}

\end{figure}

Any change in the Readout Efficiency due to the vapor cell filter necessarily should lead to a change in $g_{A,B}^{(2)}$ 
as the filter cell has no effect on $p_B$. Following the increases in $RE$ seen in Fig.~\ref{fig:wrre}(b),
a similar $\approx 35\%$ increase in the value of $g_{A,B}^{(2)}$ for $\Delta_w/2\pi=$ 40 and 60~MHz is seen in Fig.~\ref{fig:g2freq} as the absolute values of $g_{A,B}^{(2)}$ increase from 95 to 130 and 100 to 135 respectively. For $\Delta_w/2\pi=20$ MHz a
smaller increase in $g_{A,B}^{(2)}$ of $\approx 15 \%$  (from 95 to 110) corresponds closely to the similar increase of $RE$ seen in Fig.~\ref{fig:wrre}(b) as well.

The difference in the performance of the vapor cell filter for the three write detunings can be partially
 explained by effective optical depth of the atomic ensemble at different Write beam detunings.
 Based upon $\Delta_w/2\pi = 20$~MHz ($3.3 \Gamma$) and a measured
OD$_{res} = 12$ for the atomic ensemble, one would expect an effective optical depth of the cloud for the created photons to be
OD$_{3.3\Gamma} \approx~0.24$ from the equation: OD$_{\Delta} = $ OD$_{res} (1+(2\Delta/\Gamma)^2)^{-1}$~\cite{1367-2630-15-8-085027}. However that assumes
the emitted single photons interact with the entire ensemble. The probability for excitation is proportional to the atomic density 
which is largest in the center of the ensemble.  As a consequence, the average single photon
would only see half the atoms in the cloud. The resulting OD ($\approx 0.12$) leads to
$\approx 11\%$ attenuation of Noise single photons emitted by the ensemble. For the dataset with $\Delta_w/2\pi =$ 40 (60)~MHz 
there is only 3$\%$ (1$\%$) attenuation of Noise single photons due to the smaller effective optical depth from the larger detuning. Since there are 
slightly more Noise photons for the larger Write detuning, the vapor cell has a relatively larger effect on the 
write probability, readout efficiency, and non-classical correlations for $\Delta_w/2\pi =$ 40 and 60~MHz. 

\section{\label{sec:con} Conclusions}
In this paper we have shown how a temperature-controlled vapor cell filled with isotopically-pure $^{85}$Rb filled with 47 Torr of Ar buffer gas increases the Readout efficiency
and the non-classical correlations between the single photons emitted from the cold $^{87}$Rb ensemble. The filter cell achieves a substantial reduction in the detection of Noise photons by frequency-selective absorption with little effect on the detection of Signal photons. Our model showed that a buffer gas in the vapor cell was required to eliminate the detrimental effect of the $^{87}$Rb contamination in commercially available isotopically-pure $^{85}$Rb vapor cells. Our experiments showed that for Write detunings of 40 and 60 MHz the measured Readout Efficiency and $g_{A,B}^{(2)}$ of our quantum memory increased by $\approx 35\%$ as the filter cell temperature was tuned from $298~K$ to $336~K$ due to the significant decrease in the detection of Noise photons. Less dramatic effects were seen for a Write detuning of 20 MHz which were partially explained by the detuning-dependent optical depth of the cold atomic ensemble.
These results are important to maximizing the operating parameters that are critical to the realization of a quantum network based on neutral atom quantum memories.  

\section*{Acknowledgments}
We would like thank  N. Solmeyer, D. Matsukevich, and A. Gorshkov for
discussions on the quantum memory and P. Kunz for discussions on the vapor
cell.
DS is an Oak Ridge Associated Universities (ORAU) postdoctoral fellow.
Research was sponsored by the Army Research Laboratory. The views and
conclusions contained in this document are those of the Authors and should
not be interpreted as representing the official policies, either expressed
or implied, of the Army Research Laboratory or the U.S. Government. The U.S.
Government is authorized to reproduce and distribute reprints for Government
purposes notwithstanding any copyright notation herein.


\end{document}